\documentclass[iop]{emulateapj}

\usepackage{color}
\usepackage[colorlinks,linkcolor=black,anchorcolor=black,citecolor=blue,hyperfootnotes=true]{hyperref}
\usepackage[section]{placeins}
\usepackage{graphicx}
\usepackage{epstopdf}

\newcommand{\beq}[1]{\begin{equation}\label{#1}}
 \newcommand{\eeq}{\end{equation}}
 \newcommand{\bea}{\begin{eqnarray}}
 \newcommand{\eea}{\end{eqnarray}}

\newcommand{\Mpc}{\mathrm{~km~s^{-1}~Mpc^{-1}}}

\def\aj{AJ}
\def\aap{A\&A}
\def\mnras{MNRAS}
\def\apjl{ApJL}
\def\jcap{JCAP}
\def\apjs{ApJS}
\def\prd{PRD}
\def\prl{PRL}
\begin{document}

\title{Model-independent way to determine the Hubble constant and the curvature from phase shift of gravitational waves with DECIGO}

\author{Tonghua Liu\altaffilmark{1}, Shuo Cao\altaffilmark{2,3$\ast$}, Marek Biesiada\altaffilmark{4}, Yilong Zhang\altaffilmark{2,3}, Jieci Wang\altaffilmark{5}  }

\affiliation{1. School of Physics and Optoelectronic Engineering, Yangtze University, Jingzhou 434023, China; \\ 2. School of Physics and Astronomy, Beijing Normal University, Beijing 100875, China; \\ 3. Institute for Frontiers in Astronomy and Astrophysics, Beijing Normal University, Beijing 102206, China; \\ 4. National Centre for Nuclear Research, Pasteura 7, PL-02-093 Warsaw, Poland;\\ 5. Department of Physics, and Collaborative Innovation Center for Quantum Effects and Applications, Hunan Normal University, Changsha 410081, China}
\email{$\ast$ Corresponding author:\\ caoshuo@bnu.edu.cn}
\email{Marek.Biesiada@ncbj.gov.pl}
\email{jcwang@hunnu.edu.cn}
\begin{abstract}
In this Letter, we propose a model-independent method to determine the Hubble constant and curvature  simultaneously taking advantage of the possibilities of future space-borne gravitational wave (GW) detector DECIGO in combination with the radio quasars as standard rulers.
Similarly to the redshift drift in the electromagnetic domain, accelerating expansion of the Universe causes a characteristic phase correction to the gravitational waveform detectable by DECIGO. Hence, one would be able to extract the Hubble parameter $H(z)$. This could be used to recover distance-redshift relation supported by the data not relying on any specific cosmological model. Assuming the FLRW metric, and using intermediate luminosity radio quasars as standard rulers one achieves an interesting opportunity to directly assess $H_0$ and $\Omega_k$ parameters.
To test this method we simulated a set of acceleration parameters achievable by future DECIGO. Based on the existing sample of 120 intermediate-luminosity radio-quasars calibrated as standard rulers, we simulated much bigger samples of such standard rulers possible to obtain with VLBI. In the case of $(N=100)$ of radio quasars, which is the size of currently available sample, the precision of cosmological parameters determined would be $\sigma_{H_0}=2.74$ $\Mpc$ and $\sigma_{\Omega_k}=0.175$. In the optimistic scenario $(N = 1000)$ achievable by VLBI, the precision of $H_{0}$ would be improved to $1\%$, which is comparable to the result of $\sigma_{H_0} =0.54$ $\Mpc$ from \emph{Planck} 2018 TT, TE, EE+lowE+lensing data, and the precision of $\Omega_k$ would be 0.050. Our results demonstrate that such combined analysis, possible in the future, could be helpful to solve the current cosmological issues concerning the Hubble tension and cosmic curvature tension.

\end{abstract}


\section{Introduction}

Over the past few decades, the observations of type Ia supernova (SN Ia) revealed that the expansion of the universe is accelerating  \citep{1998AJ....116.1009R,2007ApJ...659...98R}. Based on the cosmological principle (supporting homogeneous and isotropic universe) and Einstein's general theory of relativity (GR), it is generally accepted that this acceleration is caused by dark energy with negative pressure.
The simplest solution is to use the cosmological constant term $\Lambda$ as a model of dark energy driving the accelerated expansion of the universe. With addition of cold dark matter, it forms the $\Lambda$CDM model. This model has been supported by the vast majority of astronomical observations over the past few decades  \citep{2016A&A...594A..13P,1998PhRvL..80.1582C,2014MNRAS.441...24A,2018PhRvD..98d3526A}. However, with the accumulation of precise astrophysical observations, some problems gradually emerged. For instance, there is a $4.4\sigma$ tension between the Hubble constant ($H_0$) measurements inferred within the $\Lambda$CDM model from cosmic microwave background (CMB) anisotropy (both temperature and polarization) data \citep{2020A&A...641A...6P} and the local measurements of the Hubble constant by the \textit{Supernova $H_0$ for the Equation of State} collaboration (SH0ES) \citep{2019ApJ...876...85R}. This inconsistency may be caused by unknown systematic errors in astrophysical observations or unrevealed new physics which is significantly different from $\Lambda$CDM \citep{2017NatAs...1E.169F,2021APh...13102605D}.
Moreover, some works suggested that besides the $H_0$ tension, spatial curvature of the universe is still an open issue \citep{2021APh...13102605D,2020NatAs...4..196D,2021PhRvD.103d1301H}.  More specifically, combining the {\it Planck} temperature and polarization power spectra data, \citet{2020A&A...641A...6P} claimed that closed universe ($\Omega_k=-0.044^{+0.018}_{-0.015}$) was supported by pure CMB data. However, the combination of {\it Planck} lensing data and low redshift baryon acoustic oscillations (BAO), revealed that flat universe with $\Omega_k=0.0007\pm0.0019$ was preferred. It should be emphasized that both methods depend, to some extent, on a particular cosmological model assumed, i.e. the non-flat $\Lambda$CDM. To better understand inconsistency between the Hubble constant and the curvature of universe inferred from local measurements and {\it Planck} observations, it is necessary to seek a completely model-independent method that can constrain both $H_0$ and $\Omega_k$ simultaneously \citep{2019PhRvL.123w1101C,2020ApJ...897..127W,2021MNRAS.503.2179Q,2022arXiv220407365L,2023JCAP...07..059L}.

Traditionally, if the Hubble parameter $H(z)$ and the luminosity distance $D_L(z)$ or angular diameter distance $D_A(z)$ are known at the same redshift $z$, one can directly measure $H_0$ and $\Omega_k$ simultaneously. The difficulty, however, is that possibilities of measuring Hubble parameters in the electromagnetic window are limited. Current cosmic chronometer sample providing such measurements via differential ages of passively evolving galaxies consists of only of 31 $H(z)$ data. Fortunately, gravitational wave (GW) observations provide such a chance to obtain $H(z)$ directly. Recently, the discovery of the first GW event, GW150914 \citep{Abbott16}, caused a great excitement in astronomy and physics communities. GW signals from inspiraling and merging compact binaries enable absolute measurements of the luminosity distance $D_L(z)$ \citep{Schutz1986}, hence they are often referred to as standard sirens. Subsequent detection of the GW170817 event \citep{Abbott17}  powered by the merger of binary neutron stars (NSs) was accompanied by electromagnetic counterpart detected in gamma-rays, X-rays, in the optical and radio bands.  If the redshifts to standard sirens can be independently determined from their electromagnetic (EM) counterparts, such GW signals can be used for cosmology. More importantly, if the adiabatic inspiral phase could be monitored long enough, the accelerating expansion of the universe would cause an additional phase shift (more details in Sec. 2.1) in the gravitational waveform, which would allow us to measure the cosmic acceleration directly \citep{Yagi2012}. This idea is similar to the redshift drift \citep{1962ApJ...136..319S,1998ApJ...499L.111L}  but requires shorter cadence of observations. Current LIGO-Virgo-Kagra (LVK) network of detectors has negligible possibility to detect cosmic acceleration because chirp signals in their frequency range are present from seconds to a couple of minutes. The future generation of space-borne  DECi-hertz Interferometer Gravitational-wave Observatory (DECIGO) will be a game changer, because the inspiral signals, ultimately to be detected by LVK ground based detectors, could be discovered and monitored a few years before the final coalescence. Thus, a cosmological model-independent way to provide $H(z)$ with covering a wide redshift range in GW domain is possible \citep{Seto2001}.

Inspired by above mentioned ideas, we will determine the curvature and the Hubble constant simultaneously using the Hubble expansion rate from simulated future space-borne detector DECIGO in combination with the ``angular size-distance" relation of ultra-compact structure in radio quasars (QSOs) from the very-long-baseline interferometry (VLBI) observations. The advantage of our work consists in assuming only the homogeneity and isotropy of the universe, thus, it is a geometric and cosmological model-independent way for the measurement of Hubble constant and spatial curvature  simultaneously. The paper is organized as follows: In Sec. 2, we present the methodology and simulated data. The results and discussion are given in Sec.~3. Finally, we summarize our findings in Sec.~4.

\section{Methodology and simulation of GW and QSO data.}

Assuming that the universe is homogeneous and isotropic on large scales, the space-time of our universe can be described by Friedmann-Lema{\^ i}tre-Robertson-Walker (FLRW) metric \citep{2008cosm.book.....W}:
\begin{equation}
ds^2 = - c^2dt^2 + a^2(t) \left[ \frac{dr^2}{1-kr^2}+r^2(d\theta^2+\sin^2\theta
d\phi^2) \right],
\end{equation}
where $t$ is the cosmic time and $r$, $\theta$, $\phi$ are the comoving spatial coordinates. The scale factor $a(t)$ is the only gravitational degree of freedom and its evolution is determined by matter and energy content in the universe. The dimensionless curvature $k=-1, 0 ,+1$ corresponds to open, flat and closed universes, respectively.
With such a metric, the angular diameter distance $D_A(z)$ can be expressed as \citep{2008cosm.book.....W}
\begin{equation}\label{eq2}
{D_A}(z) = \left\lbrace \begin{array}{lll}
\frac{D_H}{(1+z)\sqrt{|\Omega_{\rm k}|}}S_k \left(\sqrt{|\Omega_{\rm k}|}\int_{0}^{z}\frac{dz'}{E(z')}\right),\\
\frac{D_H}{(1+z)}S_k \left(\int_{0}^{z}\frac{dz'}{E(z')}\right), \\
\frac{D_H}{(1+z)\sqrt{|\Omega_{\rm k}|}}S_k \left(\sqrt{|\Omega_{\rm k}|}\int_{0}^{z}\frac{dz'}{E(z')}\right),\\
\end{array} \right.
\end{equation}
where $D_H=c/H_0$ is known as the Hubble distance, the dimensionless Hubble parameter $E(z)$ is defined as $H(z)/H_0$, where $H(z)$ is the universe expansion rate at redshift $z$ and $H_0=H(z=0)$ is the Hubble constant. The curvature parameter $\Omega_k$ is related to $k$ as
$\Omega_k=-c^2 k/(a_0H_0)^2$, where $c$ is the speed of light. For convenience, we denoted $S_k(x)=\sin(x), x, \sinh(x)$ for $\Omega_k<0, =0, >0$. If the angular diameter distance $D_A(z)$ and Hubble parameter $H(z)$ are known, then we can constrain both $H_0$ and $\Omega_k$. Standard procedure is to assume the cosmological model like $\Lambda$CDM or wCDM (i.e. the cold dark matter model with dark energy not being exactly cosmological constant) and use $H(z)$ formula suggested by the Friedman equation to calculate theoretical value of angular luminosity distance $D_{A}^{th}(z;\mathbf{p})$ at the distance corresponding to the standard ruler, whose observed distance $D_A^{obs}(z)$ can be measured. Then the chi-square based likelihood is maximized (or chi-square minimized) to obtain the parameters $\mathbf{p}$, in such case $\mathbf{p}$ comprise also cosmological model parameters like $\Omega_m, \Omega_{\Lambda}, etc.$. In this work, we use the opportunity to measure the acceleration parameter $X(z)$ (more details see Sec.2.1, Eq.~(\ref{X}) and thereafter) from an additional phase shift in the gravitational waveform based on the expected performance of the future DECIGO space-based detector. This would enable to obtain directly the Hubble parameter $H(z)$ at the redshift of NS-NS binary monitored with DECIGO.
Having the large sample of acceleration parameters $X(z_i),\; i=1,...,N$ we use Artificial Neural Network (ANN) \citep{2020ApJS..246...13W} to reconstruct the $X(z)$ function supported only by the data, without assuming any particular cosmological model.
This reconstructed relation (with the corresponding uncertainty band) can be used to calculate the distance Eq. (\ref{eq2}) to every standard ruler used to measure the distance. In this case the set of cosmological parameters $\mathbf{p}$ comprise only $H_0$ and $\Omega_k$. In the role of standard rulers, we will use the ultra-compact radio quasars. Detailed description of GW data and radio quasars is given below.

\subsection{The acceleration parameter from gravitational waveform detected by DECIGO}

In an expanding Universe, infinitesimal time intervals $dt_o$ measured in the observers frame are related to the analogous time intervals $dt_e$ in source (emitter) frame by the following (well known) relation: $\frac{dt_o}{dt_e} = \frac{a(t_o)}{a(t_e)} = \frac{1}{a(t_e)} = 1+z.$ This relation underlies the notion of the cosmological redshift. The next order derivative reads: $\frac{d^2t_o}{dt_e^2} = (1+z)^2 \left[ {\dot a} (t_o) - {\dot a}(t_e) \right] = (1+z) \left[ (1+z) H_0 - H(z) \right]$ and underlies the phenomenon of the redshift drift. Its order of magnitude is the inverse of the age of the Universe, hence this correction is usually being neglected in most cosmological considerations.
In the context of GW signals from inspiralling binaries, the relevant argument of the wave strain $h(t)$ is the time to coalescence $\Delta t = t_c - t$ with $t_c$ denoting coalescence time, measured in the observers frame. One can relate it to the time to coalescence $\Delta t_e$ measured in the source frame by: $\Delta t = \Delta T + X(z) \Delta T^2$, where $\Delta T = \Delta t_e (1+z)$ is the usual first order correction due to cosmological time dilation and the second order (Taylor expansion) factor has been denoted $X(z)$ and is called the acceleration parameter (more details in the text following Eq.(\ref{X})).
Calculating Fourier transformed ${\tilde h}(f)$ and using stationary phase approximation \citep{Yagi2012} it is straightforward to see that accelerating expansion introduces the phase shift $\Psi_{acc} = - 2 \pi f X(z) \Delta T^2$, which is additional with respect to the first order treatment neglecting the acceleration. \footnote{One should not be mislead that $\Psi_{acc}$ is simply proportional to $f$ due to frequency dependence of $\Delta T$. Rigorous dependence on $f$ is revealed in Eq.~(4).} See Fig.~1 of \citet{Kawamura2019} for an artistic illustration of this effect, i.e., GW signals coming from a neutron star binary at a given redshift with and without the acceleration of the expansion of the Universe. It was demonstrated \citep{Seto2001,Nishizawa2012} that in a gravitational wave signal emitted by a neutron star binary at the redshift of $z=1$, accelerated expansion of the Universe would cause a phase delay of about l sec during the 10-year observing time.

Waveform templates used in matched-filtering technique do not account for acceleration. Hence, with $\sim 10^7$ orbital cycles monitored during 10 years the accumulated phase shift due to acceleration would be detectable. As a second-generation space-based GW detector, DECIGO \citep{Kawamura2011} is designed to improve the detection sensitivity of GW at lower frequencies, with its most sensitive frequencies between 0.1 and 10 Hz. It can be achieved by the proposed four clusters of spacecrafts, each of which having triangular 1000 km Fabry-Perot Michelson interferometer. This design improves the determination of GW sources' position in the sky, as well as the detection sensitivity of GW at lower frequencies.
The expected sensitivity of DECIGO is $\sim 4 \times 10^{-24}$ Hz$^{-1/2}$ for one cluster, which enables the early detection of inspiraling sources, and more importantly, the direct measurement of the acceleration of the expansion of the universe.
The angular resolution is $\sim1$ arcsec$^{2}$, which enables the unique identification of the host galaxy of the binary, and improves the precision of distance measurements \citep{Seto2001}.
Therefore, DECIGO will provide us a new perspective on studying acceleration of cosmic expansion and on gravitational wave cosmology \citep{zhang2022·}.

Besides, benefiting from the precise time prediction of coalescence time and high resolution of sky location of GW sources, redshift determination from EM follow-up observations would be more reliable and frequent.
DECIGO is expected to observe $10^{6}$ GW events coming from the neutron star binaries up to the redshift $z\sim5$ \citep{Kawamura2019,Nishizawa2012}.  For our purpose, we follow the work of \citet{Zhang2022}, where they assumed that the GW event randomly occurs in any one of the galaxies, and considering the large number of spectroscopic and photometric redshift measurements in future galaxy surveys (such as JDEM/WFIRST \citep{Gehrels2010, Levi2011}, BigBOSS \citep{Schlegel2010}/DESI \citep{DESI,DESI DR} or 4MOST \citep{4MOST}), they concluded that 10,000 GW events from DECIGO with redshift determinations could be obtained by EM follow-up observations.
Considering the GW signal from NS-NS binary system whose components have masses $m_1$, $m_2$, one can define the redshifted chirp mass $\mathcal{M}_z\equiv M (1+z) \eta^{3/5}$, where $M=m_1+m_2$ is the total mass of the binary system, $\eta= m_1m_2/M^2$ is the symmetric mass ratio, and $z$ is the source redshift.
In the frequency domain, gravitational waveform including the effects of the cosmic acceleration can be written as
\begin{equation}
\tilde{h}(f)_\mathrm{acc}=\tilde{h}(f) e^{i \Psi_{\mathrm{acc}}(f)},
\end{equation}
where
\begin{equation}
\Psi_{\mathrm{acc}}(f)=-\Psi_{N}(f) \frac{25}{768} X\left(z\right) \mathcal{M}_{z} x^{-4},
\end{equation}
is the phase correction due to the cosmic acceleration, where $x\equiv (\pi \mathcal{M}_z f)^{2/3}$ and $\Psi_N(f) \equiv \frac{3}{128}(\pi \mathcal{M}_z f)^{-5/3}$.
Terms with $x^{-n}$ represent the $n$ - th Post Newtonian (PN) order, hence $\Psi_{\mathrm{acc}}(f)$ is of ``4 - PN'' order and becomes more important at lower frequencies.
The acceleration parameter $X(z)$  is defined as \citep{Seto2001,Takahashi2005}
\begin{equation} \label{X}
X(z) \equiv \frac{H_0}{2} \left(1-\frac{H(z)}{(1+z) H_0} \right).
\end{equation}
Note such a parameter can be rewritten as $X(z)=[\dot{a}(0)-\dot{a}(z)] / 2$. Thus it directly reflects whether the universe expansion has accelerated ($X(z)>0$) or decelerated ($X(z)<0$).
Angular diameter distance, given by Eq.~(\ref{eq2}) can be expressed in terms of the acceleration parameter in the following form:
\begin{equation}\label{eq2new}
{D_A}(z) = \left\lbrace \begin{array}{lll}
\frac{D_H}{(1+z)\sqrt{|\Omega_{\rm k}|}}f\left(\sqrt{|\Omega_{\rm k}|} \frac{H_0}{1+z} \int_{0}^{z}\frac{dz'}{H_0 - 2X(z')}\right),\\
\frac{c}{(1+z)^2}f\left(\int_{0}^{z}\frac{dz'}{H_0 - 2X(z')}\right), \\
\frac{D_H}{(1+z)\sqrt{|\Omega_{\rm k}|}}f\left(\sqrt{|\Omega_{\rm k}|}\frac{H_0}{1+z} \int_{0}^{z}\frac{dz'}{H_0 - 2X(z')}\right).\\
\end{array} \right.
\end{equation}

The gravitational waveform without cosmic acceleration is given by
\begin{equation}
\tilde{h}(f) =\frac{\sqrt{3}}{2}\mathcal{A}f^{-7/6}e^{i\Psi (f)} \left[ \frac{5}{4}A_{\mathrm{pol},\alpha}(t(f)) \right] e^{-i \left( \varphi_{\mathrm{pol},\alpha}+\varphi_D \right) },\label{waveform}
\end{equation}
where the amplitude
\begin{equation}
\mathcal{A}=\frac{1}{\sqrt{30}\pi^{2/3}}\frac{{\mathcal{M}_z}^{5/6}}{D_L},
\end{equation}
includes the luminosity distance $D_{L}$, which indicates that it can be measured directly. 
For the phase $\Psi(f)$, we use the restricted-2PN \citep{Kidder93} waveform including spin-orbit coupling at the 1.5PN order, which is given by
\begin{eqnarray}\nonumber
\Psi(f)&=& 2 \pi f t_{c}-\phi_{c}-\frac{\pi}{4}+\frac{3}{128}(\pi \mathcal{M} f)^{-5 / 3}\\ \nonumber
 &\times&\left[1-\frac{5}{84} \mathcal{S}^{2} \bar{\omega} x^{-1}-\frac{128}{3} \beta_{g} \eta^{2 / 5} x\right.
-\frac{2355}{1462} I_{e} x^{-19 / 6} \\ \nonumber
&+&\left(\frac{3715}{756}+\frac{55}{9} \eta\right) x
-4(4 \pi-\beta) x^{3 / 2}\\
&+&\left(\frac{15293365}{508032}+\frac{27145}{504} \eta\right.
\left.\left.+\frac{3085}{72} \eta^{2}-10 \sigma\right) x^{2}\right],
\end{eqnarray}
where the direction of orbital angular momentum and the direction of the source from the Sun are also taken into account \citep{Yagi2010}. The polarisation amplitude $A_{\mathrm{pol},\alpha}(t)$ and the polarisation phase $\varphi_{\mathrm{pol},\alpha}(t)$ represent the waveform measured by each detector ($\alpha=\mathrm{I}, \mathrm{II}$). The Doppler phase $\varphi_{D}(t)$ denotes the difference between the phase of the wave front at the detector and the phase of the wave front at the solar system barycenter. All  these quantities are explicitly given in \citet{Yagi2010}.

The waveform $\tilde{h}(f)_\mathrm{acc}$ in Eq.~(3) depends on 11 parameters:
\begin{equation}
\theta^{i}=\left(\ln \mathcal{M}_{z}, \ln \eta, \beta, t_{c}, \phi_{c}, \bar{\theta}_{\mathrm{S}}, \bar{\phi}_{\mathrm{S}}, \bar{\theta}_{\mathrm{L}}, \bar{\phi}_{\mathrm{L}}, D_{L}, X\right),
\end{equation}
where $t_{c}$ and $\phi_c$ represent the time of and the phase at the final coalescence, $\beta$ is the spin-orbit coupling parameter.
$(\bar{\theta}_{\mathrm{S}}, \bar{\phi}_{\mathrm{S}})$ indicates the direction of the source in the solar system barycenter frame, and $(\bar{\theta}_{\mathrm{L}}, \bar{\phi}_{\mathrm{L}})$ specifies the direction of the orbital angular momentum. For simplicity, we set $m_1=m_2= 1.4M_{\odot}$ and take $t_c=\phi_c=\beta=0$. For each fiducial redshift $z$, we randomly generate $10^4$ sets of $(\bar{\theta}_{\mathrm{S}},\bar{\phi}_{\mathrm{S}},\bar{\theta}_{\mathrm{L}},\bar{\phi}_{\mathrm{L}})$, and for each set, we calculate the uncertainties of the acceleration parameter $X$ estimated, using the Fisher matrix formalism:
\begin{equation}\label{eq5}
\Gamma_{ij}=4Re\int_{f_{\mathrm{min}}}^{f_{\mathrm{max}}}\frac{\partial_i\widetilde{h}^{*}(f)\partial_j\widetilde{h}(f)}{S_h(f)}df,
\end{equation}
where partial derivatives are with respect to the parameters and
$S_h(f)$ is the DECIGO noise power spectrum, which analytical expression is \citep{Kawamura2006,Kawamura2019,Yagi2011}:
\begin{eqnarray}\nonumber
S_h(f)&=&6.53\times 10^{-49} \big[1+(\frac{f}{7.36Hz})^{2} \big]+4.45\times 10^{-51} \\ \nonumber
&\times & (\frac{f}{1Hz})^{-4}\times \frac{1}{1+(\frac{f}{7.36Hz})^{2}}  \\
&+&4.94\times10^{-52}\times (\frac{f}{1Hz})^{-4}  \rm{Hz^{-1}}.
\end{eqnarray}
The lower cut-off of frequency in Eq.~(10) $f_{\mathrm{min}}=(256/5)^{-3/8} \pi^{-1} \mathcal{M}_z^{-5/8} \Delta T_{obs}^{-3/8}$ is related to the observation duration $\Delta T_{obs}$. The upper cut-off frequency of the detector $f_\mathrm{max}$ is set equal to 100 Hz. In addition, since there will be eight uncorrelated interferometric signals from DECIGO, the Fisher matrix above should be multiplied by 8 \citep{Kawamura2011,Yagi2011}.
One can use Fisher analysis to estimate the measurement accuracy of the  acceleration parameter $X$ by marginalizing other parameters in the Fisher matrix
\begin{equation}\label{eq7}
\sigma_{X} = 8^{-1/2} \left[\left( \Gamma^{-1}_{jj} \right)^{1/2}\right],
\end{equation}
where the $\sigma_{X}$ is the root-mean-square uncertainty for the $j$-th parameter (the acceleration parameter $X$), and
the $\Gamma^{-1}_{jj}$ is the covariance matrix element for
parameter $X$ calculated from the Eq.~(10).

\begin{figure}
\begin{center}
\includegraphics[scale=0.4]{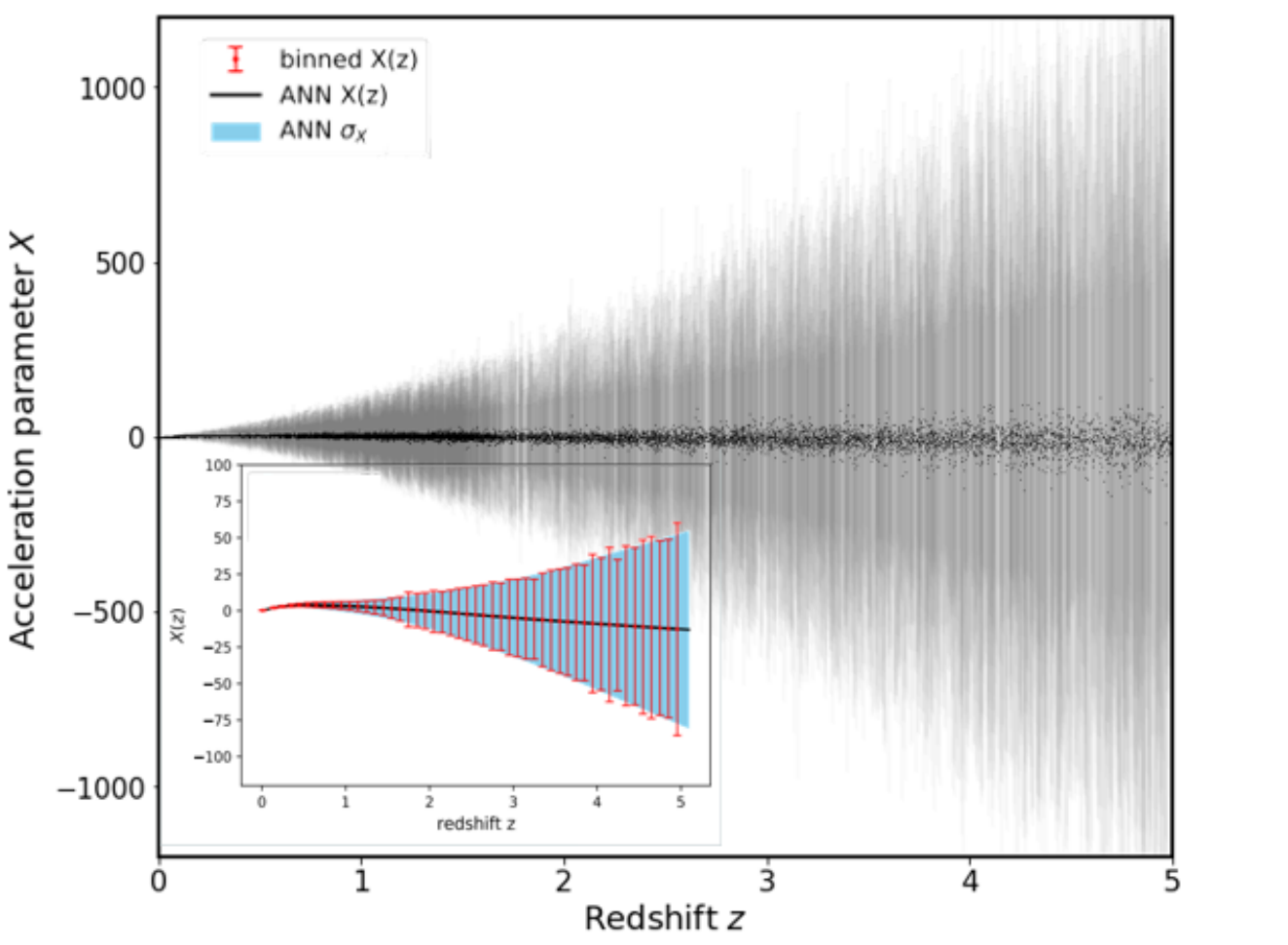}
\end{center}
\caption{Simulated acceleration parameters $X(z)$ based on the future DECIGO and the corresponding uncertainties $\sigma_X$ are shown in grey. Inset figure shows the effect of binning. Red bars represent weighted mean of $X(z)$ in redshift bins of width $\Delta z = 0.1$ with the corresponding uncertainties. Black solid line with corresponding blue uncertainty band display the ANN reconstructed $X(z)$ relation. The advantage of binning is noticeable.}
\label{fig1}
\end{figure}

Finally, the redshifts of GW sources are sampled according to the merger rate of double compact objects that reflects the star formation history
\citep{Dominik2013}, which is called ``rest frame rates" in the cosmological scenario\footnote{http:www.syntheticuniverse.org}. In our simulation, we assume flat $\Lambda$CDM as the proposed fiducial model with the cosmological parameters derived by \emph{Planck 2018} measurements \citep{2020A&A...641A...6P}.
From these simulated results ANN technique \citep{2020ApJS..246...13W} was used to reconstruct the $X(z)$ function and the corresponding uncertainty band in the redshift range $z \in [0,5]$.
Machine learning methods have become very popular in astrophysics and cosmology.  Motivated by successful applications of ANN \citep{Liu2021, Wang2020a, Qi2023}, we chose this method, which is a general method that could reconstruct a function from any kind of data without assuming any particular parameterization of the function. Moreover, this non-parametric technique does not assume Gaussianity of measured random variables, hence it is a completely data driven approach. Based on the future generation of space-borne GW detector DECIGO, we simulated a dataset of $10^4$ measurements of the acceleration parameter $X(z)$.  However, the acceleration parameters have large uncertainties, because there are too many unknown parameters in the waveform $\tilde{h}(f)_\mathrm{acc}$, which leads to large uncertainties in the estimation from the Fisher matrix. Therefore, we divided this sample into 50 bins and adopted ANN to reconstruct the acceleration parameter as a function of redshift.
We trained the ANN network on the simulated $X(z)$ data, and then predicted the $X(z)$ at other redshifts (where there were no GW data available). In general, the artificial neural network consists of an input layer, one or more hidden layers and an output layer. In this work, the input of the neural network is the redshift $z$, while the output is the corresponding cosmic acceleration parameter $X(z)$ and its respective uncertainty $ \sigma_{X}$ at that redshift. More details of data reconstruction using ANN can be found in \citep{Wang2020a,Liu2021,Qi2023}. Our strategy of non-parametric reconstruction of $X(z)$ function implies that one does not need to match the QSOs and NS-NS sources by redshift, but rather for each QSO the corresponding value of acceleration parameter will be reconstructed from the total sample of GW signals. The values of the acceleration parameters $X(z)$ and their corresponding uncertainties are shown in Fig. 1. Let us remark that individual acceleration parameters have considerable uncertainties, hence we binned the data in the redshift bins of width of width $\Delta z = 0.1$ and in each redshift bin we calculated the weighted mean, $X(z)=\frac{\Sigma_{i=1}^{50}(X_i(z)/\sigma^2_{X_i(z)})}{\Sigma_{i=1}^{50} 1/\sigma^2_{X_i(z)}}$, $\sigma^2_{{X}(z)}=\frac{1}{\Sigma_{i=1}^{50} 1/\sigma^2_{X_i(z)}}$. The ANN network was subsequently trained on these $n=50$ bins.
The final uncertainties level is similar as in the recent work \citep{Sun2023}, which used convolutional neural network (CNN) to derive the cosmic acceleration parameters.

\subsection{Angular diameter distances from radio quasars}\label{qso}

Quasars, as the brightest sources in the universe, have still a great potential to be used as useful cosmological probes, despite of  high observed dispersion and extreme variability. Observations of these objects have long been sought to study cosmology beyond the limits of supernovae. With the advances in observational technology over recent decades, quasars can be observed at multiple wave-bands, and great efforts have been made to use them as standard candles or standard rulers in cosmology, using the Baldwin effect \citep{1977ApJ...214..679B}, the Broad Line Region radius-luminosity relation \citep{2011ApJ...740L..49W}, the properties of highly accreting quasars \citep{2013PhRvL.110h1301W}, etc. Recently, \citet{Risaliti18} attempted to use quasars as standard candles through the non-linear relation between their intrinsic UV and X-ray emission. However, the large dispersion still remains the greatest problem in calibrating the QSOs as standard candles.  In this work, we focus on the ``angular size-distance'' relation of ultra-compact structure in radio quasars both observed so far and those expected from the future VLBI observations.

\begin{figure}
\begin{center}
\includegraphics[scale=0.55]{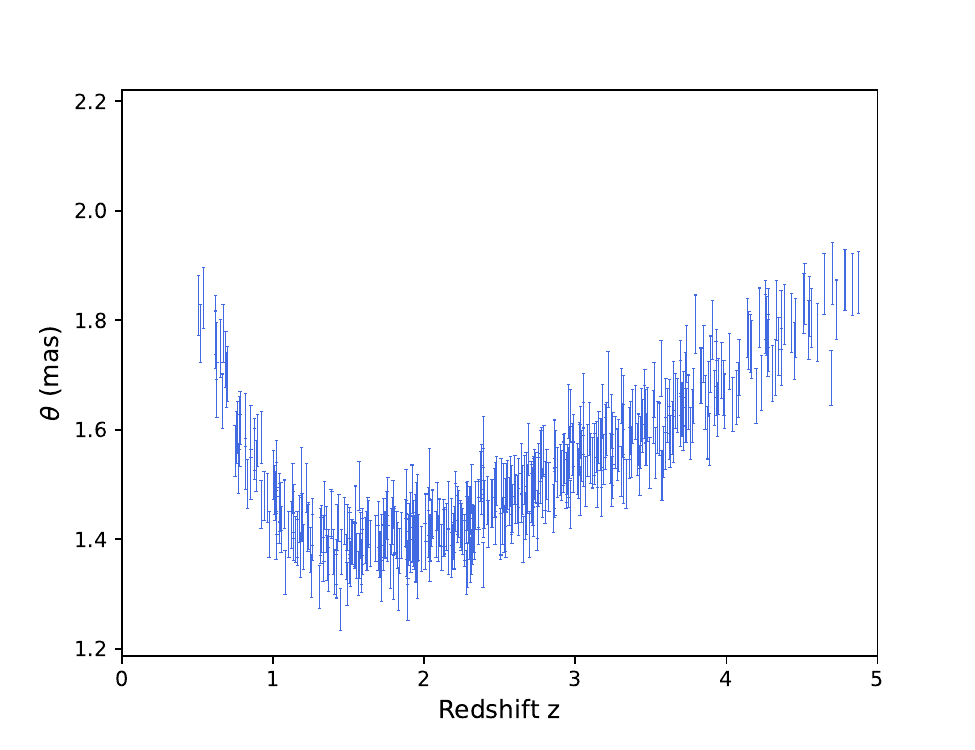}
\end{center}
\caption{Simulated sample of $1000$ angular size measurements of ultra-compact structure in radio quasars from the future VLBI observations.}
\label{fig2}
\end{figure}

With the signal received at multiple radio telescopes across the Earth's surface, together with the registered correlated intensities considering the different arrival times at various facilities, the characteristic angular size of a distant radio quasar is defined as \citep{1994ApJ...425..442G}
\begin{equation}
\theta={2\sqrt{-\ln\Gamma \ln 2} \over \pi B}, \label{thetaG}
\end{equation}
where $B$ represents the interferometer baseline measured in wavelengths, and $\Gamma=S_c/S_t$ is the visibility modulus which is defined by the ratio between the total and correlated flux densities.
In a more detailed study involving the compact structure of radio quasars,  \citet{1994ApJ...425..442G} showed that the dispersion in linear size is greatly mitigated by
retaining only the sources with a flat spectrum
($-0.38<\alpha<0.18$). Subsequently, the application of compact radio sources to cosmology has made a breakthrough \citep{1997MNRAS.285..806J,2001CQGra..18.1159V,2002ApJ...566...15L}.  With gradually refined selection technique and the elimination of possible systematic errors, \citet{Cao:2017ivt} compiled a sample of 120  intermediate-luminosity quasars ($10^{27}W/Hz < L < 10^{28} W/Hz$) selected from 613 milliarcsecond ultra-compact radio sources covering the redshift range $0.0035<z<3.787$ in 2.29 GHz VLBI survey. More importantly, they demonstrated that intermediate-luminosity compact radio quasars can be used to obtain constraints on cosmological parameters. The details regarding the angular sizes $\theta(z)$ and redshifts $z$ of 120 compact radio sources are listed in Table 1 of \citet{Cao:2017ivt}, which extends the Hubble diagram to the redshift range $0.46<z<2.76$. The angular size of the compact structure in radio QSOs can be written as
\begin{equation} \label{angular size}
\theta(z)=\frac{l_m}{D_A(z;\mathbf{p})},
\end{equation}
where $D_A(z)$ is the angular diameter distance at redshift $z$, and cosmological model parameters $\mathbf{p}$ contain the Hubble constant $H_0$ and other parameters entering the the dimensionless expansion rate expansion rate $E(z)$. The parameter $l_m$ denotes the linear size scaling factor describing the apparent distribution of radio brightness within the core, which needs to be calibrated with external indicators such as SN Ia or cosmic chronometers. In this analysis, we adopt the calibrated value of this cosmological standard ruler at 2.29 GHz as $l_m=11.03\pm0.25$ pc, through a cosmological model-independent technique based on cosmic chronometer measurements \citep{2017JCAP...02..012C}. This sample of VLBI ultra-compact structure in radio quasars, which was extended to multi-frequency analysis \citep{Cao2018EPJC}, has been widely applied in various cosmological applications at high-redshifts  \citep{Li17,ZXG,Qi17,Qi:2018atg,
Xu2018JCAP,LiuT20b,LiuT21a,2023PhLB..83837687L}.

\begin{figure}
\begin{center}
\includegraphics[scale=0.8]{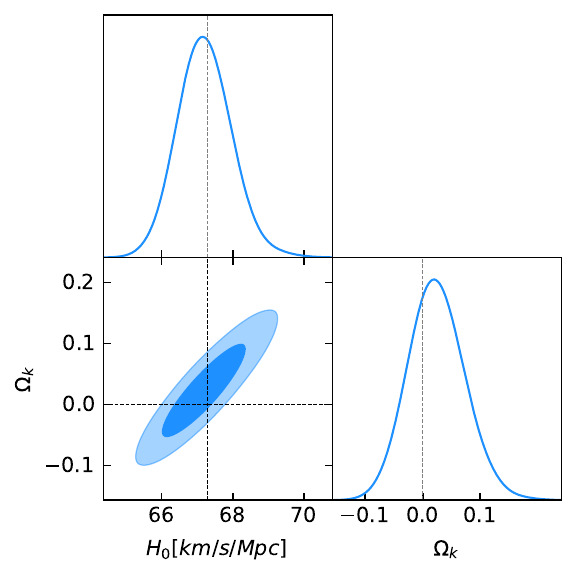}
\end{center}
\caption{The 2-D plots and 1-D marginalized distributions with 1-$\sigma$ and 2-$\sigma$ contours of cosmological parameters ($H_0$ and $\Omega_k$)  from the simulated QSO and GW data. Dashed lines represent the  fiducial parameters that we used in our simulation, i.e. $H_0=67.4\;\rm{km~s^{-1}Mpc^{-1}}$ and $\Omega_k=0$. }
\label{fig3}
\end{figure}

Considering that currently no other sample of calibrated radio quasars is available, we further simulate the possible future data on compact radio quasars from VLBI surveys based on better uv-coverage.
Our simulation is based on the flat $\Lambda$CDM with $H_0=67.4\;\rm{km~s^{-1}Mpc^{-1}}$ and $\Omega_m=0.308$ adopted from the \emph{Planck} results \citep{2020A&A...641A...6P}, i.e. the same \textbf{$\Lambda$CDM} model as used for GW simulations. Taking the linear size of intermediate-luminosity quasars as $l_m=11.03\pm 0.25$ pc and following the redshift distribution of QSOs from \citep{2016A&A...589C...2P}, we simulate 1000 ``angular size-redshift'' data representative of compact radio sources in the redshift range $0.50<z<5.00$. The fractional statistical uncertainty of the angular size $\theta$ was taken at a level of $3\%$, which is resonable for both current and future VLBI surveys based on better uv-coverage \citep{2015MNRAS.452.4274P}. Considering the intrinsic variance in the size of the AGN cores, derived from high angular resolution observations, an additional 10\% systematical uncertainty in the observed angular sizes was also assumed in our analysis. In addition, we assume that the measured angular sizes follow a Gaussian distribution $\theta_{\mathrm{means}}=\mathcal{N (\theta_{\mathrm{fid}},\sigma_{\theta})}$, where $\theta_\mathrm{fid}$ is obtained from Eq.~(\ref{angular size}) under assumed fiducial cosmological model. Simulated data are shown in Fig.~2. It should be remarked that sources with $z<0.5$ are ignored, because the era of quasar formation has ended, and the nature of quasars has changed dramatically. This problem is further expanded and elaborated in \citep{1994ApJ...425..442G}, which indicates that only the high redshift portion of the radio quasars can be used as a standard rulers. More details of the specific procedure of QSO simulation can be found in \citet{Cao2019PDU,Qi:2018atg,2019MNRAS.483.1104Q}.

\section{Results and discussion}

In order to obtain the best-fit values of $H_0$ and $\Omega_{k}$ with corresponding uncertainties, we used the Markov Chain Monte Carlo (MCMC) method \citep{Foreman2013} to minimize the $\chi^{2}$ objective function:
\begin{equation}
\chi^2=\sum \limits_{i=1}^{1000}\frac{\left[D_{A,i}^{th}(z;\mathbf{p}) - {D_{A,i}^{obs}(z)}\right]^{2}}
{\sigma_{D_{A,i}^{th}}^{2}+\sigma_{D_{A,i}^{obs}}^{2}}\;,
\end{equation}
where the $\textbf{p}$ represents the Hubble constant ${H}_0$ and cosmic curvature $\Omega_{k}$, the observed angular diameter distance is inferred from angular size-distance relation of ultra-compact structure in QSOs $D_{A,i}^{obs}(z)=l_m/\theta_i(z)$. Theoretical counterpart, $D_{A,i}^{th}(z;\mathbf{p})$ is derived from Eq.~(\ref{eq2new}) where the ANN reconstructed $X(z)$ function was used to calculate respective integrals  applying a simple trapezoidal rule. The uncertainties $\sigma_{D_{A}^{th}}^{2}$ and $\sigma_{D_{A}^{obs}}^{2}$ were calculated through the standard uncertainty propagation formula.

Combining the simulated QSO data from future VLBI observations together with the acceleration parameters from DECIGO registered GW events, we obtained the best-fit values of $H_0=67.19^{+0.77}_{-0.74}$$\Mpc$ and $\Omega_k=0.022^{+0.051}_{-0.046}$ at 68.3\% confidence level. The posterior joint distribution function along with marginal distributions is shown in Fig.~3. Our results are consistent with the fiducial values of $H_0=67.4$ $\Mpc$ and $\Omega_k = 0$ assumed, which proves the robustness of our simulation. Compared with the previous model-independent works based on the cosmic chronometer measurements $H(z)$ and distances from popular cosmological probes, our results have a higher precision than the result of combined Pantheon SN Ia plus $H(z)$ giving $\Omega_k=0.63\pm 0.34$ \citep{Wang2020a}, and the result of UV+X-ray quasars plus $H(z)$ yielding $\Omega_k=-0.92\pm 0.43$ \citep{Wei2020}. More importantly, such combination of radio and GW windows may achieve constraints with much higher precision on $H_0$, namely with $1\%$ uncertainty. Moreover, it creates an opportunity to estimate the Hubble constant and cosmic curvature using the objects at much higher redshifts ($z\sim 5$) than currently available. Therefore the future measurements of this kind would shed light on the Hubble tension, which is recognized as an issue of local vs. CMB probes.

The above mentioned result was obtained assuming the optimistic scenario of having $N=1000$ intermediate luminosity radio quasars calibrated as standard rulers. Therefore, we investigated the uncertainties of best fitted parameters as a function of QSO sample size $N$. Let us stress that the bottleneck is the QSO sample, while DECIGO is expected to provide enough data to extract acceleration parameters allowing for ANN reconstruction of $X(z)$ relation. The results are displayed in Fig. 4-5, in which two cases are considered separately. The first is the case of $H_0$ and $\Omega_k$ treated as free parameters. The second one is the precision of one parameter with a fixed prior on another one. Regarding the first case, one can see that even in the most conservative case ($N=50$), the result of $\sigma_{H_0}=4.36$ $\Mpc$ i.e. $6\%$ precision is encouraging for using the proposed cosmological model-independent method. When the number of standard rulers increases to $N=300$,  the precision of $\sigma_{H_0}=1.36$ $\Mpc$ is comparable to the result of $\sigma_{H_0}=1.3$ $\Mpc$  given by the SH0ES collaboration \citep{2019ApJ...876...85R}. If the sample size increases to $N=1000$, the precision on $H_{0}$ will be improved to $\sigma_{H_0}=0.77$, which is comparable with the result of $\sigma_{H_0} =0.54$ $\Mpc$ from \emph{Planck} 2018 TT, TE, EE+lowE+lensing data \citep{2020A&A...641A...6P}.

\begin{figure}
\begin{center}
\includegraphics[width=8.1cm,height=6cm]{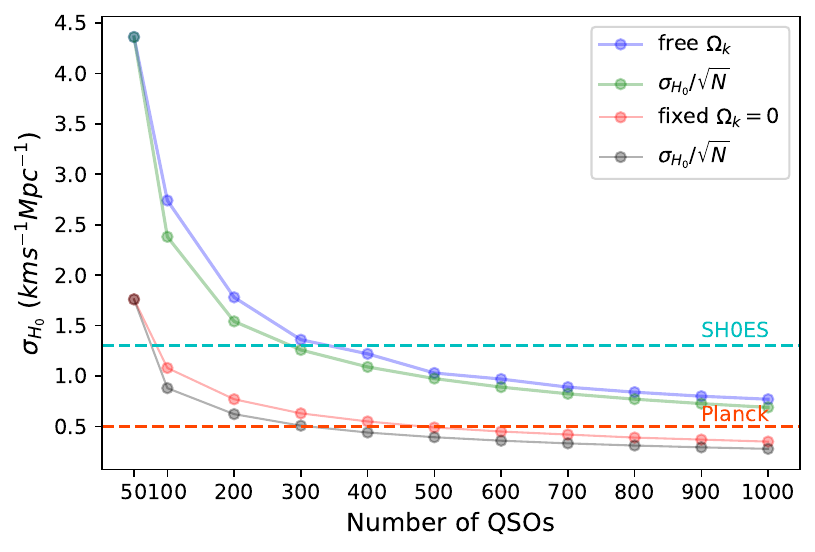}
\end{center}
\caption{The precision on $H_0$ as a function of the number of simulated QSOs.  The blue polyline represents the case of free $\Omega_k$ parameter. The red polyline represents the case of fixed $\Omega_k=0$ parameter.
For the illustration of the uncertainty floor observed, standard statistical $1/\sqrt{N}$ curves have been overplotted.
}
\label{fig4}
\end{figure}
The joint fit of the $\Omega_k$ parameter, in the optimistic scenario ($N=1000$) has the uncertainty of $\sigma_{\Omega_k}=0.050$. This is not competitive to the recent results from \emph{Planck} lensing data, where a flat universe was precisely constrained with a $1\sigma$ uncertainty of $\sigma_{\Omega_k} = 0.002$ \citep{2020A&A...641A...6P}. However, some recent works \citep{2021APh...13102605D,2020NatAs...4..196D,2021PhRvD.103d1301H} focused on the Hubble tension and other tensions in $\Lambda$CDM cosmology have raised the issue of the value of a spatial curvature of the Universe.
More specifically, combining the \emph{Planck} temperature and polarization power spectra data, these works showed that a closed universe ($\Omega_k=-0.044^{+0.018}_{-0.015}$) was supported by CMB data, which provided an explanation of anomalous high lensing amplitude \citep{2020NatAs...4..196D} reported in \emph{Planck} Legacy 2018.
The robust conclusion of these works is to emphasize the importance of alternative and cosmological model- independent methods to constrain both $H_0$ and $\Omega_k$. Our method is one of this kind and the results are encouraging. The precision would be increased if instead of two-parameter fit, one would fit a single parameter with the fixed prior on another one, which will be discussed below in more details.

\begin{figure}
\begin{center}
\includegraphics[width=8.1cm,height=5.8cm]{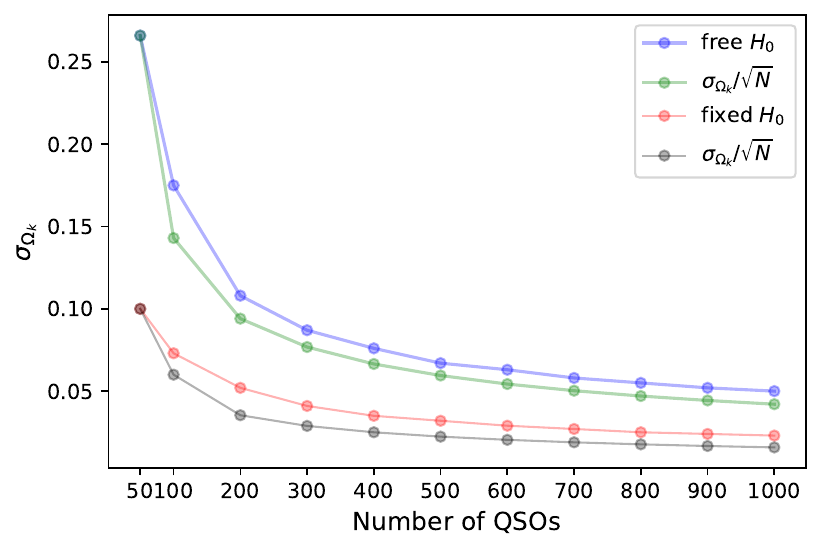}
\end{center}
\caption{The precision on $\Omega_k$ as a function of the the number of the simulated QSOs.  The blue polyline represents the free $H_0$ parameter case. The red polyline represents the fixed $H_0=67.4$  $\rm{km~s^{-1}Mpc^{-1}}$ parameter case.  For the illustration of the uncertainty floor observed, standard statistical $1/\sqrt{N}$ curves have been overplotted. }
\label{fig5}
\end{figure}
Assuming the flat Universe, as shown by the red solid polyline in Fig. 4, the constraint on $H_0$ will be improved significantly. In this case, the constraint from only $N=100$ QSOs would have uncertainty of $\sigma_{H_0}=1.08 ~\rm km\ s^{-1}\ Mpc^{-1}$, which is slightly better than the result of $\sigma_{H_0}=1.3 ~\rm km\ s^{-1}\ Mpc^{-1}$ reported by the SH0ES collaboration. Only $N=500$ QSOs are required to produce the results comparable with those obtained by \emph{Planck} 2018 TT, TE, EE+lowE+lensing data. In the most optimistic scenario ($N=1000$), we can get a result of $\sigma_{H_0}=0.35$ $\rm km\ s^{-1}\ Mpc^{-1}$, which is at $0.5\%$ level. All these results demonstrate that the combination of GW+QSO data proposed by us could become an important  cosmological probe, precise enough to address the Hubble tension issue. With the fixed prior of $H_0=67.4~\rm km\ s^{-1}\ Mpc^{-1}$, the constraints on $\Omega_k$ are also considerably improved. This can be understood in terms of degeneracy between $\Omega_k$ and $H_0$, which can also be seen from the joint constraint confidence regions of $H_0$ and $\Omega_k$ shown in Fig.3. In the most conservative case ($N=50$) the uncertainty of $\Omega_k$ is $\sigma_{\Omega_k}=0.100$, while in the most optimistic scenario ($N=1000$), $\sigma_{\Omega_k}=0.023$. This is not as good as the result from the \emph{Planck} 2018 data \citep{2020A&A...641A...6P}, yet it refers to an alternative measurement using extragalactic objects in more local Universe than CMB. An interesting point is that as the number of QSOs increases, the constraint on $\Omega_k$ does not improve significantly after reaching $N=500$. From the statistical point of view, when the number of observations increases by $N$, the precision should increase by $\sqrt{N}$. The saturation observed suggests the existence of a feature in the analysis that prevents from going lower.
We have checked how the hypothetical assumption of zero uncertainty for the angular diameter distance influence the uncertainty of cosmological parameters inferred. It turned out that it has negligible effect $\sigma_{H_0}$ and $\sigma_{\Omega_k}$ were only slightly different from the above mentioned values (at the second significant digit).
Better understanding of the origin of such an uncertainty floor would be desired and is left for future studies.
Moreover, in future applications of our method with the real data, the influence of systematic effects cannot be ignored and should be studied carefully.

Our method is built on two pillars: GWs as a source of $H(z)$ data and QSOs as source of $D_A(z)$ data.
In order to gain some understanding of relative importance of uncertainties related to GW signals and QSO, we did the following test. Namely, based on $N=1000$ QSO data paired with GW reconstructed $X(z)$ values, we left the uncertainty of QSO data unchanged but doubled the uncertainty of GW inference. The results we obtained in this case were: $\sigma_{H_0}=1.18$ $\Mpc$ and $\sigma_{\Omega_k}=0.070$. Similarly, we kept the uncertainty of GW unchanged but doubled the uncertainty from QSOs. We obtained the precision of Hubble constant to be $\sigma_{H_0} = 1.04$ $\Mpc$ and the precision of curvature parameter to be $\sigma_{\Omega_k} = 0.079$. Fig. 6 displays the $1\sigma$ and $2\sigma$ confidence level contours for these two cases for estimated parameters. In conclusion, we found that the contribution of QSO and GW data uncertainties to the precision of parameter constraints were comparable.

\begin{figure}
\begin{center}
\includegraphics[scale=0.83]{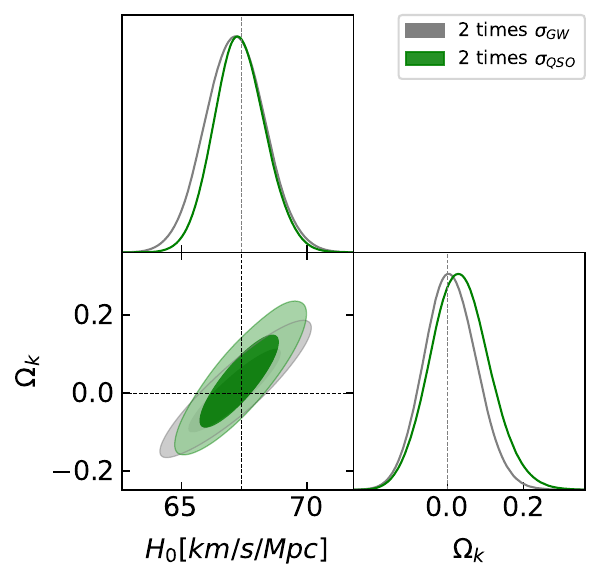}
\end{center}
\caption{1D and 2D marginalized probability distributions of $H_{0}$ and $\Omega_{k}$ as in Fig.3 but corresponding to doubling the uncertainties from GW or QSO data respectively, while keeping the other unchanged.
Dashed lines represent the  fiducial parameters that we used in our simulation, i.e. $H_0=67.4\;\rm{km~s^{-1}Mpc^{-1}}$ and $\Omega_k=0$. }
\label{fig6}
\end{figure}

Finally, we tested the influence of the GW data on the fits. For this purpose we simulated $N=1000$ GW sources, reconstructed corresponding $X(z)$ function and analysed jointly with $N=1000$ QSOs. The result was $H_0=65.69 \pm 1.71$ $\Mpc$ and $\Omega_k=0.03 \pm 0.142$ at 68.3\% confidence level. One can see that with an order of magnitude smaller sample of GW signals the uncertainties raised $2-3$ times relative to the main case studied. The precision would be at the present level attainable by local cosmological probes. Hence, the GW signal sample size and quality are crucial.

\section{Conclusion}
In this Letter, we proposed a cosmological model-independent method to determine the Hubble constant and curvature parameter simultaneously based on the future generation of space-borne GW detector DECIGO in  combination with the data regarding intermediate-luminosity radio-quasars calibrated as standard rulers obtained from the VLBI observations. Since the accelerating expansion of the Universe would cause an additional phase shift in the gravitational waveform, we can extract the Hubble parameter $H(z)$ useful for further  cosmological studies. On the other hand, the future VLBI surveys will observe a large amount of compact structures in radio quasars, which would provide the angular diameter distances $D_A(z)$ based on a simple ``angular size-distance'' relation.  Such combination of observations performed in the radio and GW windows  creates a valuable opportunity to directly test $H_0$ and $\Omega_k$ at much higher precision.

To fully demonstrate the potential of such combination of GW+QSO data-set, we simulated respective data representative of the expected yield of the DECIGO and VLBI. Simulation of the future VLBI data on QSOs was based on the properties of currently available sample of $N=120$ intermediate-luminosity radio-quasars calibrated as standard rulers \citep{Cao:2017ivt}. The same fiducial cosmology was assumed in both GW and QSO simulations. In the most optimistic case of $N=1000$ QSO calibrated as standard rulers, joint fits for $H_0$ and $\Omega_k$ were consistent with the input simulation values and their $68.3\%$ confidence level uncertainties were $\sigma_{H_0} = 0.75 $ $\Mpc$ and $\sigma_{\Omega_k} = 0.048$, respectively.
It means that achievable precision is at the level of $1\%$. Besides the joint fit on two parameters simultaneously, we considered fits of one parameter with fixed priors on the second one. In such cases precision improved as could be expected. In particular the precision of $H_0$ with a fixed prior on $\Omega_k$ can reach $0.5\%$ in the most optimistic case of $N=1000$ QSOs. We also discussed the performance of the method on QSO samples of different size (from $N=50$ to $N=1000$). In particular regarding $\Omega_k$ fit the uncertainty behavior did not reflect simple $1/\sqrt{N}$ behaviour, but saturated at $N=500.$
This phenomenon remains to be fully understood, but urges a careful consideration of systematic effects in future applications of the proposed method to the real data.

As a final remark, there are many potential ways to improve our results. In the future, multi-frequency VLBI observations will yield more high-quality quasar observations based on better uv coverage \citep{2015MNRAS.452.4274P}. Radio quasars having a compact structure, will be measured with higher angular resolution, and lower statistical and systematic uncertainty. On the other hand, we also look forward to a large amount of future data, not only from the radio quasars, but also from other astronomical probes, such as  SNe Ia and BAO, allowing us to further improve the precision of $H_0$ and $\Omega_k$ measurements. The results we obtained are very encouraging regarding the method proposed, which is alternative and complementary to CMB analysis. In light of current tensions in cosmology when local probes suggest different values of important parameters like the Hubble constant, from those obtained by CMB analysis, every cosmological probe alternative to already existing is valuable.

\acknowledgments
The authors are grateful the referee for careful reading the manuscript and numerous insightful comments. Liu T.-H. was supported by National Natural Science Foundation of China under Grant No. 12203009; Chutian Scholars Program in Hubei Province (X2023007). Cao S. was supported by Beijing Natural Science Foundation No. 1242021; National Natural Science Foundation of China under Grants Nos. 12021003, 11633001, and 11920101003; the Strategic Priority Research Program of the Chinese Academy of Sciences, Grant No. XDB23000000; and the Interdiscipline Research Funds of Beijing Normal University. Biesiada M. was supported by Hubei Province Foreign Expert Project (2023DJC040). Wang J. C. was supported by the National Natural Science Foundation of China under Grant No. 12122504 and 12035005.


\label{lastpage}

\end{document}